\documentclass[a4paper]{article}

\usepackage{algorithm, algorithmic,amsmath,amssymb,authblk,framed,graphicx,multirow}
\usepackage{longtable}
\usepackage{booktabs}
\usepackage[switch,pagewise]{lineno}
\usepackage[usenames]{color}
\usepackage{url}
\usepackage{supertabular}
\usepackage[figuresright]{rotating}
\usepackage{array}
\usepackage{lscape}
\usepackage{subfigure}

\newcommand{\PreserveBackslash}[1]{\let\temp=\\#1\let\\=\temp}
\newcolumntype{C}[1]{>{\PreserveBackslash\centering}p{#1}}
\newcolumntype{R}[1]{>{\PreserveBackslash\raggedleft}p{#1}}
\newcolumntype{L}[1]{>{\PreserveBackslash\raggedright}p{#1}}

\usepackage{mathptmx}      
\usepackage{latexsym}
\usepackage[misc]{ifsym}

\begin{document}

\setlength{\rotFPtop}{0pt plus 1fil}

\title{Algorithms for the minimum sum coloring problem: a review}

\author[1,2]{Yan Jin\thanks{yanjin.china@hotmail.com}}
\author[2]{Jean-Philippe Hamiez\thanks{jean-philippe.hamiez@univ-angers.fr}}
\author[2,3]{Jin-Kao Hao\thanks{jin-kao.hao@univ-angers.fr (Corresponding author)}}
\affil[1]{School of Computer Science and Technology, Huazhong University of Science and Technology, No. 1037, Luoyu Road, Wuhan, China}
\affil[2]{LERIA, Universit\'{e} d'Angers, 2 Bd. Lavoisier, 49045 Angers, France}
\affil[3]{Institut Universitaire de France, 1 Rue Descartes, 75231 Paris, France}

\date{\textit{Artificial Intelligence Review, DOI: 10.1007/s10462-016-9485-7, 2016}}

 	
\maketitle

\begin{abstract}
The Minimum Sum Coloring Problem (MSCP) is a variant of the well-known vertex coloring problem which has a number of AI related applications. Due to its theoretical and practical relevance, MSCP attracts increasing attention. The only existing review on the problem dates back to 2004 and mainly covers the history of MSCP and theoretical developments on specific graphs. In recent years, the field has witnessed significant progresses on approximation algorithms and practical solution algorithms. The purpose of this review is to provide a comprehensive inspection of the most recent and representative MSCP algorithms. To be informative, we identify the general framework followed by practical solution algorithms and the key ingredients that make them successful. By classifying the main search strategies and putting forward the critical elements of the reviewed methods, we wish to encourage future development of more powerful methods and motivate new applications.

\end{abstract}

\section{Introduction}
\label{sec_intro}

Given a graph $G$, a proper $k$-coloring of $G$ is an assignment of $k$ different colors $\{1, \ldots, k\}$ to the vertices of $G$ such that two adjacent vertices receive two different colors. The classical graph vertex coloring problem (GCP) is to find a proper (or legal) $k$-coloring with the minimum number of colors $\chi(G)$ (i.e., the chromatic number of $G$) for a general graph $G$. The minimum sum coloring problem (MSCP) is a variant of the GCP and aims to determine a proper $k$-coloring while minimizing the sum of the colors assigned to the vertices. MSCP was proposed by Kubicka \cite{Kubicka1989} in the field of graph theory and by Supowit \cite{Supowit1987} in the field of VLSI design. MSCP has applications in VLSI design, scheduling and resource allocation for instance \cite{Bar-Noy&al1998,Bonomo&al2015,Kroon&al1996,Malafiejski2004,Sen&al1992}. MSCP is also related to other generalizations or variants of GCP like sum multi-coloring \cite{Bar-Noy&al1999}, sum list coloring \cite{Berliner&al2006} and bandwidth coloring \cite{Dam2008}.

Like the classical vertex coloring problem, MSCP is notable for its practical applicability and theoretical intractability. Indeed, in the general case, the decision version of MSCP is NP-complete \cite{Kroon&al1996,Kubicka1989} and approximating the minimum color sum within an additive constant factor is NP-hard \cite{Kubicka&al1991}. As a result, MSCP is a computationally challenging problem and any algorithm able to determine the optimal solution of the problem is expected to require an exponential complexity. Due to its high computational complexity, polynomial-time algorithms exist only for some special cases of the problem (see Section \ref{sec_approximation}) and solving the problem in the general case remains an imposing challenge.

In the past several decades, much effort has been devoted to developing various approximation algorithms and practical solution algorithms. Approximation algorithms aim to provide solutions of provable quality while practical solution algorithms try to find sub-optimal solutions as good as possible within a bounded and acceptable computation time. The class of heuristic and metaheuristic algorithms has been mainly developed since 2009 and has enlarged our capacity of finding improved solutions on the benchmark graphs. Representative examples of the existing heuristic algorithms include greedy algorithms \cite{Li&al2009,Moukrim&al2010}, tabu search \cite{Bouziri&Jouini2010}, breakout local search \cite{Benlic&Hao2012}, iterated local search \cite{Helmar&Chiarandini2011}, ant colony \cite{Douiri&Elbernoussi2012}, genetic and memetic algorithms \cite{Douiri&Elbernoussi2011,Jin&Hao2016,Jin&al2014,Kokosinski&Kwarciany2007,Moukrim&al2013,Wang&al2013} as well as heuristics based on independent set extraction \cite{Wu&Hao2012,Wu&Hao2013}.

To the best of our knowledge, there is only one review published one decade ago in 2004 \cite{Kubicka2004} that focuses on polynomial-time algorithms for specific graphs, MSCP generalizations (or variants) and applications. For the purpose of solving MSCP, the first studies essentially concerned the development of approximation algorithms and simple greedy algorithms. Research on practical solution algorithms of MSCP was relatively new and appeared around 2009. Nevertheless, important progresses have been made since that time. The purpose of this paper is thus to provide a comprehensive review of the most recent and representative MSCP algorithms. To be informative, we identify the general framework followed by the existing heuristic and metaheuristic algorithms and their key ingredients that make them successful. By classifying the main search strategies and putting forward the critical elements of the reviewed methods, we wish to encourage future development of more powerful methods and motivate new applications.

In the following sections, we first provide a general definition of  MSCP, then a brief introduction of approximation algorithms in Section \ref{sec_approximation}, followed by the presentation of the studied heuristics and metaheuristics in Section \ref{sec_approach}. Section \ref{sec_upper&lower} presents lower and upper bounds. Before concluding, Section \ref{sec_results} introduces  MSCP benchmark instances and summarizes the computational results reported by the best performing algorithms on these instances.

\section{Definitions and formulation of MSCP}
\label{sec_Definitions}

Let $G=(V,E)$ be a simple undirected graph with vertex set $V={\{v_1,\ldots,v_n\}}$ and edge set $E \subset V \times V$.
A proper $k$-coloring $c$ of $G$ is a mapping $c: V \rightarrow {\{1, \ldots, k\}}$ such that $c(v_i) \neq c(v_j)$, $\forall {\{v_i, v_j\}} \in E$. Equivalently, a proper $k$-coloring can be defined as a partition of $V$ into $k$ mutually disjoint independent sets (or color classes) $V_1, \ldots, V_k$ such that $\forall u, v \in V_i$ $ (i=1, \ldots, k), {\{u, v\}} \notin E$.  The objective of  MSCP is to find a proper $k$-coloring $c$ with a minimum sum of the colors that are assigned to the vertices of $V$. The minimum sum of colors for MSCP is called the \textit{chromatic sum} of $G$, and is denoted by $\sum(G)$. The \textit{strength} $s(G)$ of a graph $G$ is the smallest number of colors over all optimal sum colorings of $G$. Obviously, the chromatic number $\chi(G)$ of $G$ from the classical vertex coloring problem is a lower bound of $s(G)$, i.e., $\chi(G) \leq s(G)$.

Let $\mathcal{C}(G)$ be the set of all proper $k$-coloring of $G$ and the minimization objective $f(c)$ ($c\in\mathcal{C}(G)$) of MSCP is given by Eq. (\ref{eq1MSCP}).

\begin{equation}\label{eq1MSCP}
f(c) = \sum_{i=1}^n c(v_i) \ \textrm{or} \
f(c) = \sum_{l=1}^k l|V_l|
\end{equation}
where $|V_l|$ is the cardinality of $V_l$ and $|V_1| \geq \ldots \geq |V_k|$ with the chromatic sum given by:
\begin{equation}\label{eq1}
\sum(G) = \min_{c \in \mathcal{C}(G)} f(c)
\end{equation}

\begin{figure}[h]
\begin{center}
\begin{tabular}{c@{\qquad\qquad}c}
\includegraphics[scale=0.85]{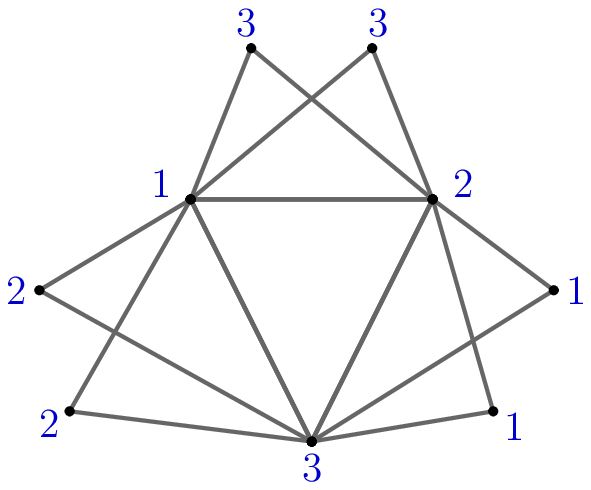} & \includegraphics[scale=0.85]{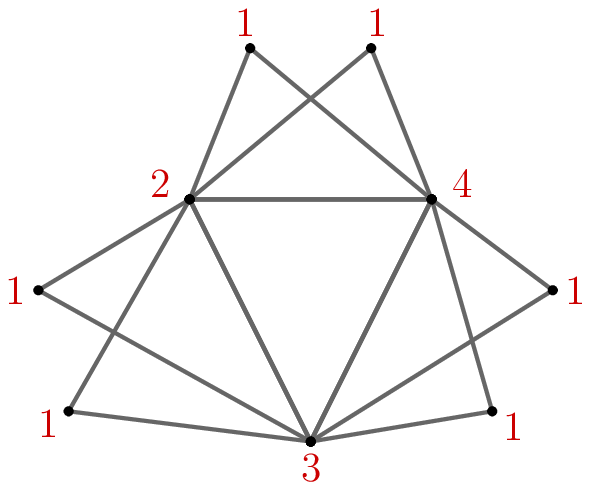}\\
\end{tabular}
\caption{An illustrative example for MSCP \cite{Jin&Hao2016}. The optimal coloring of the graph leads to an upper bound of the chromatic sum of the graph.}
\label{fig_example}
\end{center}
\end{figure}

Figure \ref{fig_example} shows an illustrative example for MSCP. The graph has a chromatic number $\chi(G)$ of 3 (left figure), but requires 4 colors to achieve the chromatic sum (right figure). Indeed, with the given 4-coloring, we achieve the chromatic sum of 15 while the 3-coloring of left figure leads to a suboptimal sum of 18 (upper bound).

As shown in \cite{Sen&al1992}, MSCP can be conveniently formulated as an integer linear programming problem as follows:

\begin{equation}
\label{eq2}
\begin{array}{ll}
\textrm{minimize} & g(x)=\sum_{i=1}^n \sum_{l=1}^k l \cdot x_{il}\\
\textrm{subject to} &
\left\{
\begin{array}{l}
\sum_{l=1}^k x_{il} = 1, i \in \{{1, \ldots, n}\}  \\
x_{il} + x_{jl} \leq 1, \forall \{v_i, v_j\} \in E, l \in \{{1,\ldots,k}\} \\
x_{il} \in \{{0, 1}\}
\end{array}
\right.
\end{array}
\end{equation}

where $x_{il} = 1$ ($i \in \{1,\ldots,n\}, l \in \{1,\ldots,k\}$) if $v_i$ is assigned color $l$, $x_{il} = 0$ otherwise.

The first constraint of this ILP model ensures that each vertex receives a single color while the second constraint states that two adjacent vertices cannot be assigned the same color. This linear model can be solved by any ILP solver like CPLEX \cite{Wang&al2013}. Finally, as shown in \cite{Wang&al2013}, MSCP can also be formulated as a binary quadratic programming model.

\section{Polynomial-time and $k$-approximation algorithms for MSCP}
\label{sec_approximation}

One notes that till now no exact algorithm especially designed for MSCP was reported in the literature except the general solution approach used in \cite{Wang&al2013} which applies CPLEX to the integer linear programming formulation (Eq. (\ref{eq2})). On the other hand, a number of polynomial-time and $k$-approximation algorithms have been proposed for \textit{specific} classes of graphs, such as trees, interval graphs, bipartite graphs, etc \cite{Borodin&al2012,Hajiabolhassan2000,Jiang&West1999,Kosowski2009,Malafiejski2004}. These algorithms exploit particular properties of the special graphs considered. In what follows, we briefly recall the main characteristics of these specific classes of graphs:

\begin{itemize}
	\item A \textit{cograph}, also called $P_4$-free graph, is a graph that does not contain the path $P_4$ for any four vertices\footnote{A path $P_4$ is a sequence of 4 vertices, say $\left(v_1, v_2, v_3, v_4\right)$, such that $\left\{v_i,v_{i+1}\right\} \in E$ $\forall i \in \left\{1, 2, 3\right\}$ and $\left\{v_i, v_{i+k}\right\} \notin E$ $\forall k \in \left\{1, 2, 3, 4\right\} \backslash \left\{i-1,i+1\right\}$.};
	\item \textit{$P_4$-reducible} graphs are a generalization of cographs where every vertex belongs to at most one $P_4$;
	\item \textit{$P_4$-sparse} graphs generalize $P_4$-reducible graphs by imposing that every set of five vertices induces at most one $P_4$;
	\item \textit{Unicyclic} graphs contain exactly one cycle;
	\item A \textit{partial $k$-tree} $G$ is a graph with treewidth of at most $k$, where the treewidth is the size of the largest vertex set in a tree decomposition of $G$;
	\item A graph is \textit{outerplanar} if it is planar (it can be embedded in the plane without crossing edges) and all its vertices lie on the exterior face;
	\item The \textit{line} graph $L(G)$ of any graph $G=(V, E)$ is such that its vertex set is $E$ and two vertices of $L(G)$ are adjacent if their corresponding edges in $G$ are incident;
	\item In an \textit{interval} graph, each vertex corresponds to an interval (over the set of real numbers for instance) and there is an edge between two vertices if their corresponding intervals intersect.
\end{itemize}

In the field of VLSI design, Kroon et al. \cite{Kroon&al1996} considered the ``optimum cost chromatic partition problem'' (OCCP), whose definition is similar to MSCP. For this problem, they introduced a linear-time algorithm for trees (see also \cite{Kubicka&Schwenk1989}). Other classes of graph optimally solved in linear time include cographs \cite{Jansen2000} or unicyclic graphs \cite{Kubicka2005} for instance.

In \cite{Jansen2000}, Jansen found that the OCCP can be solved in polynomial time for partial $k$-trees. Then, Salavatipour presented a polynomial-time algorithm for $P_4$-reducible graphs \cite{Salavatipour2003}. Furthermore, Bonomo and Valencia-Pabon studied $P_4$-sparse graphs and found a large sub-family of $P_4$-sparse graphs that can be solved in polynomial time \cite{Bonomo&Valencia-Pabon2014}. A cubic algorithm has also been proposed for outerplanar graphs \cite{Kubicka2005}.

Bar-Noy et al. proposed a 2-approximation algorithm\footnote{A $k$-approximation algorithm ensures to return a solution whose evaluation / cost is no more than a factor $k$ of the optimum.} for line graphs and showed a $(\Delta + 2) /3$-approximation algorithm for graphs with maximum degree $\Delta$ \cite{Bar-Noy&al1998}. Then, Bar-Noy and Kortsarz proposed a 10/9-approximation algorithm for bipartite graphs \cite{Bar-Noy&Kortsarz1998}. This approximation ratio was next improved to 27/26 by Malafiejski et al. \cite{Malafiejski&al2004} which is the best ratio for bipartite graphs to our knowledge. For interval graphs, Nicoloso et al. presented a 2-approximation algorithm \cite{Nicoloso&al1999}, the best known ratio for this class of graphs being 1.796 \cite{Halldorsson&al2003}. Let us finally mention a 2-approximation algorithm for the entire class of $P_4$-sparse graphs \cite{Bonomo&al2015}.

\section{Heuristics and metaheuristics for MSCP}
\label{sec_approach}

Since these approximability results cannot be generalized to an arbitrary graph, for practically solving MSCP in the general case, a number of heuristic and metaheuristic algorithms have been proposed recently. In this section, we review the most representative and effective MSCP heuristic and metaheuristic algorithms which belong to three large classes of methods: greedy algorithms, local search heuristics, and evolutionary algorithms. For each reviewed algorithm, we identify its key ingredients, and highlight if the search process is constrained in the feasible space or is allowed to visit infeasible regions. We also provide in Table \ref{tableAlgorithm} a summary of the reviewed algorithms as well as indicators about their performances.

\subsection{Greedy algorithms}
\label{subsec_greedyAlgorithms}

Greedy algorithms are among the first heuristics proposed for MSCP. These algorithms are generally fast, simple, and easy to implement. Nevertheless, they usually achieve results of poor quality. On the other hand, given their particular features (speed and simplicity), they can advantageously be integrated into other more elaborated approaches where the greedy heuristic is used to generate an initial solution and seeds the search process. For instance, they can be used to provide initial upper bounds for an exact algorithm or to build the initial solution(s) for local search heuristics and evolutionary algorithms.

Two families of greedy algorithms for MSCP are proposed in \cite{Li&al2009}: MDSAT(\emph{n}) and MRLF(\emph{n}). They are based on the two well-known greedy coloring heuristics DSATUR \cite{Brelaz1979} and RLF \cite{Leighton1979}.

The original DSATUR heuristic employs the saturation degree $dsat$ of a vertex\footnote{$dsat(v_i)$ is the number of colors used to color the vertices adjacent to $v_i$.} as the selection criterion to dynamically determine the next vertex to color. MDSAT(\emph{n}) improves DSATUR by considering the impact of coloring a vertex where the impact is measured based on the number of vertices whose $dsat$ would (not) be changed.
The original RLF heuristic follows the partition perspective of a vertex coloring. It colors as many non-adjacent vertices as possible with one color before going to another color. MRLF(\emph{n}) which extends RLF is based on the idea of selecting the next candidate vertex $v$ for coloring such that it reduces the chance of using a new color next and keeps the current color class as large as possible. To achieve this goal, MRLF(\emph{n}) implements sophistic greedy rules which rely on the cardinality of a subset of uncolored vertices that could be colored with and without using a new color.

A more complicated greedy heuristic (EXSCOL) is proposed in \cite{Wu&Hao2012,Wu&Hao2013}. It is based on independent set extraction and is highly effective for hard and large graphs. At each iteration, EXSCOL first identifies an independent set $S$ as large as possible by using a tabu search procedure. Secondly, it searches as many independent sets as possible of the same size $|S|$ to build a pool of candidate independent sets. Then, EXSCOL determines a maximum number of disjoint independent sets by solving a maximum set packing problem. Finally, the vertices of each extracted independent set receive the same smallest available color to form a color class. The above process is repeated until the graph becomes empty. Notice that there is no procedure to reconsider the extracted independent sets such that it is impossible for EXSCOL to attain an optimal solution once a ``bad'' independent set has been extracted.

\subsection{Local search heuristics}
\label{subsec_localSearch}

Local search (or neighborhood search) heuristics progressively modify a candidate solution $c$ by local transformations until a stop condition is reached \cite{Gendreau&Potvin2010}. The two key components of a local search procedure are the evaluation function and the move (or transformation) operator which are defined on a given search space.

The evaluation function is used to assess the quality of a given coloring. The existing MSCP algorithms employ one of two types of evaluation function according to whether feasible or infeasible colorings are visited. For algorithms that explore only feasible solutions (i.e. proper colorings), the minimization function $f$ (i.e., the sum of colors, Eq. (\ref{eq1MSCP})) of the MSCP problem is directly used. On the other hand, algorithms that visit both feasible and infeasible solutions usually call for an augmented evaluation function $f_p$ which combines the objective function $f$ and a penalty function $p$.

In local search algorithms, one iteratively uses one or more move operators to transform the incumbent solutions $c$ to generate new neighboring solutions $c'$. The set of neighboring solutions that can be reached by applying a move operator ($mv$) to the current solution forms the neighborhood (denoted by $N_{mv}$). We describe the commonly used operators as follows.

\begin{itemize}
	\item \textit{One-move} changes the color of a vertex in the current solution by moving a vertex $v$ from its current color class $V_i$ to another color class $V_j$ ($i \neq j$). This operator can generate both proper or improper colorings and thus can be used to explore feasible and infeasible regions of the coloring search space;
	\item \textit{Swap} displaces a vertex $v$ from its current color class $V_i$ to another color class $V_{j}$ (as \textit{One-move}) and then moves all adjacent vertices $u$ of $v$ to $V_i$. This operator can generate both proper or improper colorings;
	\item \textit{Exchange} swaps a subset of vertices $A \subset V_i$ ($|A|>1$) and another subset of vertices $B \subset V_j$ ($|B|>1$) ($i \neq j$) such that the subgraph induced by $A \cup B$ is a connected component \cite{Jin&al2014}. The new solution $c'$ is feasible (respectively infeasible) if the starting solution $c$ is feasible (infeasible).
\end{itemize}

In what follows, we classify the representative local search algorithms into two categories according to the adopted neighborhood(s): single neighborhood search and multi-neighborhood search. Since local search can get stuck in local optima, most local search algorithms for MSCP use some diversification techniques to help the search to escape local optima encountered during the search. This is typically achieved by applying one or more perturbation operators to change a local optimum in a random or dedicated way.

\subsubsection{Single neighborhood search}

The tabu search (TS) algorithm proposed in \cite{Bouziri&Jouini2010} adapts the tabu algorithm designed for the classic vertex coloring problem \cite{Galinier&Hao1999,Hertz1987}. It starts with a random coloring and visits both proper and improper colorings with the neighborhood $N_{One-move}$ induced by the \textit{One-move} operator. If there exist conflicting vertices, TS chooses a best move (according to its evaluation function $f_p$) to change the color of a conflicting vertex. Otherwise, TS picks a (non-conflicting) vertex and change its color at random. The above steps are repeated until a stopping criterion is satisfied. This algorithm relies simply on the tabu list for its diversification and does not call for other perturbation mechanism. This algorithm only showed limited computational results.

The breakout local search (BLS) algorithm described in \cite{Benlic&Hao2012} jointly uses two descent methods and an adaptive multi-perturbation strategy to escape local optima. The basic idea of BLS is to use descent-based local search to discover local optima and employ adaptive perturbations to continually visit different search regions in the search space. BLS explores both feasible and infeasible solutions with the help of the \textit{One-move} operator. At each iteration, if the current solution $c$ is a feasible coloring, BLS applies a first descent search procedure to attain a local optimum in terms of the objective function $f$. If $c$ is an infeasible coloring (i.e., with conflicting vertices), BLS applies another descent search procedure guided by an augmented evaluation function which takes into account both the objective function $f$ and the conflicting vertices. BLS is characterized by its adaptive perturbation strategy which, upon the discover of a local optimum, triggers dedicated perturbation operations to escape the local optimum trap. Based on the information on the search state, the perturbation strategy of BLS introduces a varying degree of diversification by dynamically determining the number of perturbation moves to be applied and by adaptively selecting the suitable moves (random or directed perturbations).

\subsubsection{Multi-neighborhood search}

The MDS(5)+LS algorithm \cite{Helmar&Chiarandini2011} applies an iterated multi-neighborhood search and also explores feasible and infeasible regions of the search space. It first employs the \textit{Swap} operator until no further improvement exists in terms of its augmented evaluation function. Note that the obtained solution is not necessarily a proper coloring. If this is the case, MDS(5)+LS switches then to the \textit{One-move} operator to repair the solution. Additional colors can be used to guarantee that the final coloring is proper at the end of this search phase. Finally, it assigns all the vertices with their smallest legal color and changes the color labels according to the sorted cardinality of the color classes $V_l$ ($|V_1| \geq \ldots \geq |V_k|$). Afterward, a random perturbation operator is applied which consists in moving some vertices from their current color class to another color class at random. This perturbed solution is then used as the starting point of the next round of the search procedure.

\begin{sidewaystable}
\begin{footnotesize}
\caption{Main heuristic and metaheuristic algorithms for MSCP}\label{Summary}
\begin{tabular}{L{2.0cm}L{2.0cm}L{2.5cm}L{3.0cm}L{2.0cm}L{6.0cm}}
\hline
Algorithm name &Reference & Type of approach & Neighborhoods & Perturbation & Comments on performance\\
\hline
MDSAT(\emph{n}) MRLF(\emph{n}) &\cite{Li&al2009}(2009) &Greedy search & - & - &A family of improved greedy algorithms based on the well-known greedy coloring strategies DSATUR and RLF.\\[2ex]
TS       & \cite{Bouziri&Jouini2010}(2010)  & Local search  & $N_{One-move}$ & No & A very simple tabu search but the results are better than those of the greedy algorithms MDSAT(\emph{n}) and MRLF(\emph{n}). \\[2ex]
MDS(5)+LS &\cite{Helmar&Chiarandini2011}(2011) & Local search & $N_{One-move}$ $\&$ $N_{Swap}$ & Yes & An iterated multi-neighborhood search combined with a random perturbation procedure achieving better results than MDSAT(\emph{n}), MRLF(\emph{n}) and TS.\\[2ex]
BLS &\cite{Benlic&Hao2012}(2012) & Local search & $N_{One-move}$ & Yes & A breakout local search combining a greedy descent strategy with an adaptive perturbation step. It performs well on the small DIMACS graphs.\\[2ex]
EXSCOL &\cite{Wu&Hao2012,Wu&Hao2013}(2012) &Greedy + tabu search & No  & No & A complicated greedy algorithm, based on independent sets extraction with tabu search, which is quite effective for large graphs.\\[2ex]
MASC &\cite{Jin&al2014}(2014) &Evolutionary search & $N_{One-move}$ \& $N_{Exchange}$  & Yes &A memetic algorithm based on a double-neighborhood tabu search and a multi-parent crossover operator. Most results are better than those of the neighborhood search heuristics.\\[2ex]
MA-MSCP  & \cite{Moukrim&al2013}(2013) &Evolutionary search & $N_{One-move}$  & Yes &A genetic algorithm with a two-parents crossover operator combined with a local search based on a hill climbing and a ``destroy \& repair'' procedures. Results are comparable to those of MASC.\\[2ex]
HESA &\cite{Jin&Hao2016}(2016) &Evolutionary search & $N_{One-move}$  & Yes &A hybrid search algorithm based on a jointly use of two crossover operators and an iterated double-phase tabu search procedure. The lower and upper bounds obtained by the HESA are highly competitive with the best known results in the literature.\\
\hline
\end{tabular}
\label{tableAlgorithm}
\end{footnotesize}
\end{sidewaystable}

\subsection{Evolutionary algorithms}
\label{subsec_evolutionaryAlgorithms}

Different from local search algorithms which are based on a single solution, evolutionary algorithms use a pool of solutions and try to find gradually better solutions by applying genetic operators (e.g., crossover, mutation, \ldots) to solutions of the population \cite{Gendreau&Potvin2010}.

The most popular evolutionary algorithms for MSCP follow the hybrid evolution framework called the memetic algorithm which jointly uses a recombination operator and a local search improvement to explore the search space \cite{Gendreau&Potvin2010}. They include, for instance, the MASC algorithm \cite{Jin&al2014}, MA-MSCP algorithm \cite{Moukrim&al2013} and the HESA hybrid search algorithm \cite{Jin&Hao2016}. Besides, an early parallel genetic algorithm PGA \cite{Kokosinski&Kwarciany2007} employs assignment and partition crossovers, first-fit mutation, and proportional selection without any local search improvement.

The MASC memetic algorithm \cite{Jin&al2014} follows the design guidelines of memetic algorithms for discrete optimization \cite{Hao2012} and combines a multi-parent crossover operator (called MGPX) and a double-neighborhood tabu search procedure. MGPX is a variant of the well-known GPX crossover originally proposed for the classical vertex coloring problem \cite{Galinier&Hao1999}. It builds the color classes of the offspring (which is always a proper coloring) one by one and transmits entire color classes as large as possible until all vertices of the offspring are colored. Besides, the tabu search procedure applies the two different and complementary neighborhoods induced by \textit{Exchange} and \textit{One-move} in a token-ring way to find good local optima (according to the objective function $f$) until the search is stagnating. MASC employs a dedicated perturbation operator to diversify the search. MASC only explores the feasible search space of MSCP.

MA-MSCP is another hybrid genetic algorithm \cite{Moukrim&al2013} that also focuses on the feasible search space. It includes a two-parent crossover operator (yet another adaptive variant of GPX \cite{Galinier&Hao1999}), a hill-climbing local search algorithm and a ``destroy \& repair'' procedures. During the local search phase, the hill-climbing procedure is first applied to improve the current solution by using the \textit{One-move} operator. To escape local optima, MA-MSCP then applies the ``destroy \& repair'' strategy, which randomly removes some vertices and re-inserts each of them into its largest available color class while keeping the solution feasible. If there is no such a color class, the vertex is moved to a new color class. MA-MSCP employs the above two procedures alternately until no further improvement can be obtained.

HESA is also a hybrid search algorithm \cite{Jin&Hao2016} that alternates between feasible and infeasible regions of the search space. HESA relies on a double-crossover recombination method and an iterated double-phase tabu search procedure. The recombination method jointly uses a diversification-guided crossover and a grouping-guided crossover to generate promising offspring solutions. During the double-phase tabu search procedure, it first checks if the given solution $c$ is a proper coloring. If $c$ is proper, the first tabu search is called to improve its sum of colors. Otherwise, another tabu search is used to attain a proper coloring which is further improved by the first tabu search to obtain a better sum of colors. The double-phase tabu search only explores the $N_\textrm{One-move}$ neighborhood. For the purpose of search diversification, HESA applies a conditional mixed perturbation strategy: 1) apply the \textit{Swap} operator to a randomly chosen vertex to transform the incumbent solution, or 2) replace the current solution by the last local optimum.

Table \ref{Summary} summarizes the reviewed existing heuristic algorithms with their main characteristics including the type of search paradigm, the neighborhood(s) and the presence or absence of a perturbation strategy together with a comment on their relative performance.

Finally, we mention the BQP-PR evolutionary algorithm \cite{Wang&al2013} which relies on a binary quadratic programming formulation of the problem (see Section \ref{sec_Definitions}) and combines a path relinking approach with a tabu search procedure.

\section{Bounds for MSCP}
\label{sec_upper&lower}

We will refer here to ``theoretical'' (lower and upper) bounds if they are formally proved, see Section \ref{subsec_theoreticalBounds}. By opposition, ``computational'' bounds introduced in Section \ref{subsec_computationalBounds} designate those obtained running \emph{approximate} algorithms.

\subsection{Theoretical bounds}
\label{subsec_theoreticalBounds}

Recall that for any undirected simple graph $G=(V,E)$ with $n = |V|$ vertices and $m = |E|$ edges, the chromatic number $\chi(G)$ is the smallest number of colors needed to color the vertices of $G$ such that a proper $k$-coloring exists and the chromatic sum $\sum(G)$ is the minimum sum of the colors assigned to all vertices among all proper $k$-colorings of $G$. In this section, we list the current known theoretical lower and upper bounds of MSCP according to \cite{Kokosinski&Kwarciany2007,Moukrim&al2013,Thomassen&al1989}.

\begin{equation}\label{eq3}
\begin{split}
 &\sum(G) \leq n + m  \\
 &\lceil {\sqrt{8m}} \rceil  \leq \sum(G) \leq \lfloor \frac{3(m+1)}{2} \rfloor\\
 &n+\frac{\chi(G)(\chi(G)-1)}{2} \leq \sum(G) \leq \lfloor \frac{n(\chi(G)+1)}{2} \rfloor
\end{split}
\end{equation}

From Eq.(\ref{eq3}), one easily observes that the best theoretical lower and upper bounds available for MSCP are respectively $LB_t = \max \{\lceil {\sqrt{8m}} \rceil, n+\frac{\chi(G)(\chi(G)-1)}{2}\}$ and $UB_t = \min \{n + m, \lfloor \frac{3(m+1)}{2} \rfloor, \lfloor \frac{n(\chi(G)+1)}{2} \rfloor\}$.

\subsection{Computational bounds}
\label{subsec_computationalBounds}

Given that MSCP is to find a proper $k$-coloring while minimizing the sum of the colors assigned to the vertices, Eq. (\ref{eq1MSCP}) gives a computational upper bound for MSCP.

Let $G' =(V, E') (E' \subseteq E)$ be any partial graph of $G=(V,E)$, $\sum(G')$ is a lower bound of $\sum(G)$ since any proper coloring of $G$ must be a proper coloring of $G'$: $\sum(G) \ge \sum(G')$.

Partial graphs considered in the literature to estimate the computational lower bound $f_{LB}$ include bipartite graphs (trees and paths) \cite{Garey&Johnson1979,Kroon&al1996} and cliques \cite{Moukrim&al2010,Wu&Hao2013}, while graph decomposition into cliques\footnote{A clique is a complete graph where all the vertices are pairwise adjacent. A clique decomposition of a graph is a partition of the vertex set $V$ into a collection of cliques.} provide better bounds according to \cite{Moukrim&al2010}. Let $c = \{S_1, S_2, $ $\ldots, S_k\}$ be a clique decomposition of $G$, then Eq. (\ref{eq4}) gives a computational lower bound for MSCP since there is a single way of coloring any clique $S_l$ (with $|S_l|$ colors) and the sum of colors of $S_l$ is $|S_l|(|S_l| + 1)/2$.

\begin{equation} \label{eq4}
f_{LB}(c) = \sum_{l=1}^k \frac{|S_l|(|S_l|+1)}{2}
\end{equation}

\begin{figure}
\begin{center}
\begin{tabular}{c@{\qquad\qquad}c}
\includegraphics[scale=0.85]{example2.eps} & \includegraphics[scale=0.85]{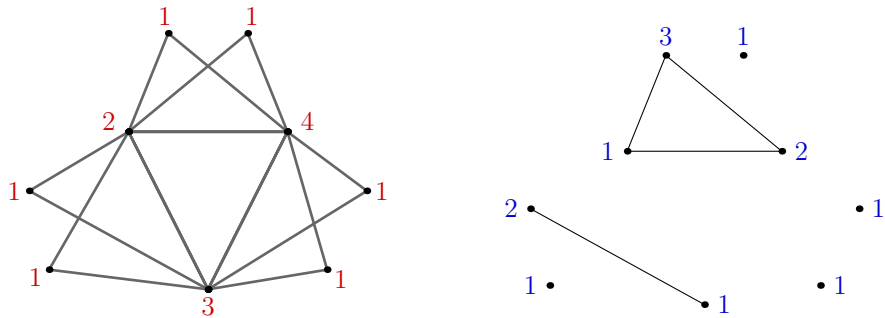}\\
\end{tabular}
\caption{An illustrative lower bound via clique decomposition. The right figure is a clique decomposition of the graph on the left.}
\label{fig_exampleLB}
\end{center}
\end{figure}

Figure \ref{fig_exampleLB} shows an illustrative lower bound via clique decomposition. We decompose $G$ into six cliques by ignoring some edges of the original graph $G$ and obtain the chromatic sum $\sum(G') = 13$ (right figure). Clearly, this is a lower bound for MSCP while the chromatic sum $\sum(G) = 15$ (left figure).

To obtain a clique decomposition, one popular approach is to find a proper coloring of the complementary graph $\bar G$ of $G$ \cite{Helmar&Chiarandini2011,Jin&Hao2016,Moukrim&al2013,Wu&Hao2013}, since each color class of $\bar G$ is a clique of $G$.

\section{Benchmark and performance evaluation}
\label{sec_results}

In this section, we first introduce a set of MSCP instances (benchmarks) that are commonly used to assess the performance of MSCP algorithms and then provide indications about the performances of the reviewed MSCP algorithms. Due to many different factors (programming languages, running platforms, experimental protocols...), it is quite difficult to draw definitive conclusions. Nevertheless, we try to provide some useful indications with respect to their performance in terms of best and average results.

\subsection{Benchmark}
\label{subsec_TestInstances}

There exists a set of 94 frequently used benchmark instances often used for performance evaluation of MSCP algorithms. 58 instances are part of the COLOR 2002--2004 competitions\footnote{\url{http://mat.gsia.cmu.edu/COLOR02}} while the remaining 36 instances come from the second DIMACS challenge\footnote{\url{http://dimacs.rutgers.edu/Challenges/}}.  Compared to the well-known DIMACS instances, the COLOR 2002-2004 instances are relatively easy except the four large ``wap'' graphs. These instances refer to various topologies and densities, which can be classified into the 13 following types:

\begin{itemize}
    \item Twelve classical random graphs (DSJC$n.d, n \in \{125, 250, 500, 1\,000\}, d \in \{1, 5, 9\}$);
    \item Three geometric graphs (DSJR500.$d$, $d \in \{1c, 1, 5\}$);
    \item Six flat graphs (flat$300\_\chi\_0$ with $\chi \in \{20, 26, 28\}$ and flat$1000\_\chi\_0$ with $\chi \in \{50, 60, 76\}$);
    \item Twelve Leighton graphs (le450\_$\chi$a, le450\_$\chi$b, le450\_$\chi$c, le450\_$\chi$d, $\chi \in \{5, 15, 25\}$);
    \item Four latin square graph (latin\_sqr\_10 and qg.order$\chi$, $\chi \in \{30, 40, 50\}$);
    \item Two very large random graphs (C2000.5 and C4000.5);
    \item Fourteen graphs based on register allocation (fpsol2.i.$a$, inithx.i.$a$, zeroin.i.$a$, mulsol.i.$b$, $a \in \{1, 2, 3\}$ and $b \in \{1, 2, 3, 4, 5\}$);
    \item Two graphs from the scheduling area (school1 and school1\_nsh);
    \item Twenty four graphs from the Donald Knuth's Stanford GraphBase (miles$n$ with $n \in \{250, 500, 750, 1000, 1500\}$, anna, david, huck, jean, homer, games120, queen8.12, and queen$a.a$, $a \in \{5, \ldots, 16\}$);
   \item Five graphs based on the Mycielski transformation (myciel$a$, $a \in \{3, 4, 5, 6, 7\}$);
   \item Four graphs that have a hard-to-find four clique embedded (mug$n\_a$, $n \in \{88, 100\}$, $a \in \{1, 25\}$);
   \item Two ``insertion'' graphs (2-Insert\_3 and 3-Insert\_3);
   \item Four graphs from real-life optical network design problems (wap05, wap06, wap07, and wap08).
\end{itemize}

\begin{table}[h]
\begin{scriptsize}
\caption{Main characteristics of MSCP benchmark (94 instances)} \label{table_benchmark}
\begin{tabular}{@{}l@{ }r@{ }r@{\quad}c@{ }r@{\quad}r@{ }r@{\qquad}l@{ }r@{ }r@{\quad}c@{ }r@{\quad}r@{ }r@{}}
\hline
Graph $G$ & $n$ & $m$  &$d$ &$\chi(G)$  &$LB_t$ &$UB_t$ & Graph $G$ & $n$ & $m$  &$d$ &$\chi(G)$ &$LB_t$ &$UB_t$ \\
\hline
myciel3&11&20&0.36 &4&17&27&zeroin.i.1&211&4100&0.19 &49&1387&4311\\
myciel4&23&71&0.28 &5&33&69&zeroin.i.2&211&3541&0.16 &30&646&3270\\
myciel5&47&236&0.22 &6&62&164&zeroin.i.3&206&3540&0.17 &30&641&3193\\
myciel6&95&755&0.17 &7&116&380&wap05&905&43081&0.11 &50&2130&23077\\
myciel7&191&2360&0.13 &8&219&859&wap06&947&43571&0.10 &40&1727&19413\\
anna&138&493&0.05 &11&193&631&wap07&1809&103368&0.06 &$\leq 41$&2629&37989\\
david&87&406&0.11 &11&142&493&wap08&1870&104176&0.06 &$\leq 42$&2731&40205\\
huck&74&301&0.11 &11&129&375&qg.order30&900&26100&0.06 &30&1335&\underline{13950}\\
jean&80&254&0.08 &10&125&334&qg.order40&1600&62400&0.05 &40&2380&\underline{32800}\\
homer&561&1628&0.01 &13&639&2189&qg.order60&3600&212400&0.03 &60&5370&\underline{109800}\\
queen5.5&25&160&0.53 &5&36&\underline{75}&DSJC125.1&125&736&0.09 &5&135&375\\
queen6.6&36&290&0.46 &7&57&144&DSJC125.5&125&3891&0.50 &17&261&1125\\
queen7.7&49&476&0.40 &7&70&\underline{196}&DSJC125.9&125&6961&0.90 &44&1071&2812\\
queen8.8&64&728&0.36 &9&100&320&DSJC250.1&250&3218&0.10 &$\leq 8$&278&1125\\
queen8.12&96&1368&0.30 &12&162&\underline{624}&DSJC250.5&250&15668&0.50 &$\leq 28$&628&3625\\
queen9.9&81&1056&0.33 &10&126&445&DSJC250.9&250&27897&0.90 &$\leq 72$&2806&9125\\
queen10.10&100&1470&0.30 &11&155&600&DSJC500.1&500&12458&0.10 &$\leq 12$&566&3250\\
queen11.11&121&1980&0.27 &11&178&726&DSJC500.5&500&62624&0.50 &$\leq 47$&1581&12000\\
queen12.12&144&2596&0.25 &12&210&936&DSJC500.9&500&112437&0.90 &$\leq 126$&8375&31750\\
queen13.13&169&3328&0.23 &13&247&1183&DSJC1000.1&1000&49629&0.10 &$\leq 20$&1190&10500\\
queen14.14&196&4186&0.22 &14&287&1470&DSJC1000.5&1000&249826&0.50 &$\leq 82$&4321&41500\\
queen15.15&225&5180&0.21 &15&330&1800&DSJC1000.9&1000&449449&0.90 &$\leq 222$&25531&111500\\
queen16.16&256&6320&0.19 &16&376&2176&DSJR500.1&500&3555&0.03 &12&566&3250\\
school1&385&19095&0.26 &14&476&2887&DSJR500.1c&500&121275&0.97 &84&3986&21250\\
school1-nsh&352&14612&0.24 &14&443&2640&DSJR500.5&500&58862&0.47 &122&7881&30750\\
miles250&128&387&0.05 &8&156&515&flat300\_20\_0&300&21375&0.48 &20&490&\underline{3150}\\
miles500&128&1170&0.14 &20&318&1298&flat300\_26\_0&300&21633&0.48 &26&625&4050\\
miles750&128&2113&0.26 &31&593&2048&flat300\_28\_0&300&21695&0.48 &28&678&4350\\
miles1000&128&3216&0.40 &42&989&2752&flat1000\_50\_0&1000&245000&0.49 &50&2225&\underline{25500}\\
miles1500&128&5198&0.64 &73&2756&4736&flat1000\_60\_0&1000&245830&0.49 &60&2770&30500\\
fpsol2.i.1&496&11654&0.09 &65&2576&12150&flat1000\_76\_0&1000&246708&0.49 &76&3850&38500\\
fpsol2.i.2&451&8691&0.09 &30&886&6990&le450\_5a&450&5714&0.06 &5&460&\underline{1350}\\
fpsol2.i.3&425&8688&0.10 &30&860&6587&le450\_5b&450&5734&0.06 &5&460&\underline{1350}\\
mug88\_1&88&146&0.04 &4&94&220&le450\_5c&450&9803&0.10 &5&460&\underline{1350}\\
mug88\_25&88&146&0.04 &4&94&220&le450\_5d&450&9757&0.10 &5&460&\underline{1350}\\
mug100\_1&100&166&0.03 &4&106&250&le450\_15a&450&8168&0.08 &15&555&3600\\
mug100\_25&100&166&0.03 &4&106&250&le450\_15b&450&8169&0.08 &15&555&3600\\
2-Insert\_3&37&72&0.11 &4&43&92&le450\_15c&450&16680&0.17 &15&555&3600\\
3-Insert\_3&56&110&0.07 &4&62&140&le450\_15d&450&16750&0.17 &15&555&3600\\
inithx.i.1&864&18707&0.05 &54&2295&19571&le450\_25a&450&8260&0.08 &25&750&5850\\
inithx.i.2&645&13979&0.07 &31&1110&10320&le450\_25b&450&8263&0.08 &25&750&5850\\
inithx.i.3&621&13969&0.07 &31&1086&9936&le450\_25c&450&17343&0.17 &25&750&5850\\
mulsol.i.1&197&3925&0.20 &49&1373&4122&le450\_25d&450&17425&0.17 &25&750&5850\\
mulsol.i.2&188&3885&0.22 &31&653&3008&latin\_sqr\_10&900&307350&0.76 &$\leq 97$&5556&44100\\
mulsol.i.3&184&3916&0.23 &31&649&2944&C2000.5&2000&999836&0.50 &$\leq 145$&12585&147000\\
mulsol.i.4&185&3946&0.23 &31&650&2960&C4000.5&4000&4000268&0.50 &$\leq 259$&37670&522000\\
mulsol.i.5&186&3973&0.23 &31&651&2976& games120&120&638&0.09 &9&156&600\\
\hline
\end{tabular}
\end{scriptsize}
\end{table}

Table \ref{table_benchmark} gives the detailed characteristics of the benchmark graphs. Columns 2--5 and 9--12 indicate the number $n$ of vertices, the number $m$ of edges, the density $d = 2m/n(n-1)$ and the chromatic number $\chi(G)$ of each graph. Columns 6--7 and 13--14 show the best theoretical lower and upper bounds of the chromatic sum ($LB_t$ and $UB_t$ respectively). Underlined entries (in all tables) indicate that theoretical upper bounds equal the computational upper bounds while no theoretical lower bound equals the computational lower bound. Note that, since the chromatic number $\chi(G)$ of some difficult graphs are still unknown, we use the minimum $k$ for which a $k$-coloring has been reported for $G$ in the literature instead of $\chi(G)$ to compute $LB_t$ and $UB_t$ using the $\min / \max$ equations introduced in Section \ref{subsec_theoreticalBounds}.

\subsection{Performance of MSCP algorithms}
\label{subsec_performance}

Based on the benchmark introduced in the previous section, Table \ref{table_many_algorithms} (see the Appendix) summarizes the computational results of six representative and effective MSCP algorithms presented in Section \ref{sec_approach}: BLS \cite{Benlic&Hao2012}, MASC \cite{Jin&al2014}, MDS(5)+LS \cite{Helmar&Chiarandini2011}, EXSCOL \cite{Wu&Hao2012,Wu&Hao2013}, MA-MSCP \cite{Moukrim&al2013} and HESA \cite{Jin&Hao2016}. Columns 1--3 present the tested graph and its best known lower and upper bounds ($f^b_{LB}$ and $f^b_{UB}$ respectively, in bold face when optimality is proved), the following 18 columns give the detailed computational results of the six algorithms. ``--'' marks for the reference algorithms mean non-available results. The results in terms of solution quality (best / average lower and upper bounds, $f^*_{LB} / f^a_{LB}$ and $f^*_{UB} / f^a_{UB}$ respectively) are directly extracted from the original papers. Computing times are not listed in the table due to the difference of experimental conditions (platforms, programming languages, stop conditions...). Nevertheless, the second and third lines of the heading respectively indicate the main computer characteristic (processor frequency) and the stop condition to have an idea of the maximum amount of search used by each approach. Note that there is no specific stop condition for EXSCOL since its extraction process ends when the current graph becomes empty. Furthermore, some heuristics can halt before reaching the stop criterion, when a known (lower) bound is reached for instance.

\begin{figure}[!htbp]
\centering
\subfigure[Lower bounds]{
\begin{minipage}[b]{0.85\textwidth}
\includegraphics[scale=0.52]{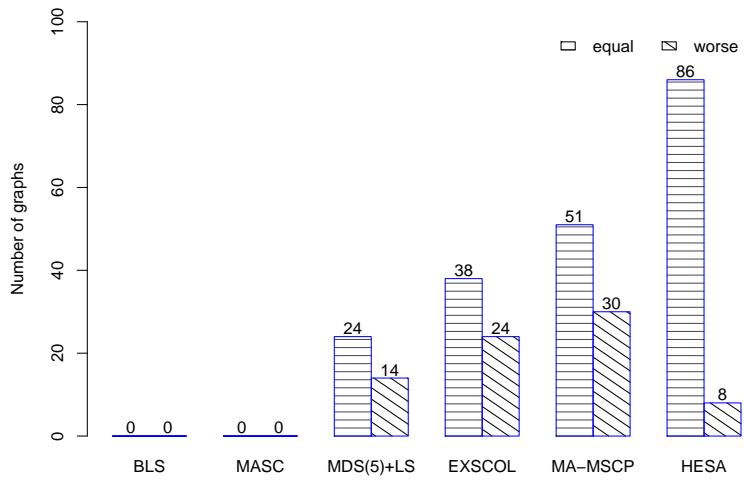}
\end{minipage}
}
\subfigure[Upper bounds]{
\begin{minipage}[b]{0.85\textwidth}
\includegraphics[scale=0.52]{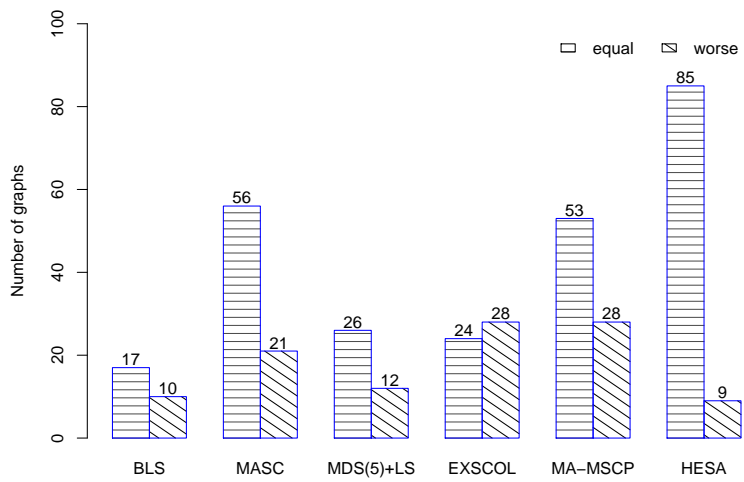}
\end{minipage}
}
\caption{The performance of six representative MSCP algorithms. The y-axis shows the number of graphs for which an algorithm attains a result equal to or worse than the best known reported bound.} \label{fig_comparisons}
\end{figure}

From Table \ref{table_many_algorithms}, one observes that only HESA reports results for all the 94 graphs of the benchmark. Besides, MDS(5)+LS, EXSCOL, MA-MSCP, and HESA provide lower and upper bounds while BLS and MASC only give an upper bound. Additionally, Figure \ref{fig_comparisons} provides performance information of each of the six algorithms compared to the best known upper and lower bounds. One observes that no algorithm can reach all the best known results. BLS and MASC attain the best upper bounds for 17 graphs out of the 27 tested graphs and for 56 graphs out of the 77 tested graphs respectively. MDS(5)+LS reaches the best lower (upper) bound for 24 (26) instances out of 38. EXSCOL reaches the best lower and upper bounds for 38 (out of 62 graphs) and 24 (out of 52 graphs) respectively. MA-MSCP reaches the best lower / upper bound for 51 / 53 graphs out of 81. HESA equals the best lower (upper) bound for 86 (85) instances out of 94.

Since the number of tested graphs differs from one algorithm to another, the performance of these algorithms cannot be compared from a statistical viewpoint. However, from Table \ref{table_many_algorithms} and Figure \ref{fig_comparisons}, we can roughly conclude that BLS, MASC, MDS(5)+LS, EXSCOL, MA-MSCP and HESA are currently the most effective algorithms for solving the MSCP problem.

From the theoretical and computational bounds reviewed above, we can make the following observations:
\begin{itemize}
\item Optimality is proved for 21 instances out of the 94 tested graphs since the best upper bounds are equal to the best lower bounds (see entries in bold in Table \ref{table_many_algorithms});
\item 12 theoretical upper bounds equal the computational upper bounds while no theoretical lower bound equals the computational lower bound (underlined in Tables \ref{table_benchmark}--\ref{table_many_algorithms});
\item The theoretical upper bounds of queen$a.a~(a \in \{11, 12, 13, 14, 15, 16\})$ are equal to the best computational lower bounds meaning optimal results;
\item Table \ref{table_many_algorithms} shows that the best computational lower bounds of some easy graphs (myciel$a$, $a \in \{3, 4, 5, 6\}$, for instance) are not equal to the optimal upper bounds (optimality proved with CPLEX \cite{Wang&al2013}). Hence, the method of decomposing the graph introduced in Section \ref{subsec_computationalBounds} is not good enough in some cases and should be improved.
\end{itemize}

\section{Perspectives and conclusion}
\label{Sec_Conclusion}

This review is dedicated to recent approximation algorithms and practical solution algorithms designed for the minimum sum coloring problem which attracted increasing attention in recent years. MSCP is a strongly constrained combinatorial optimization problem which is theoretically important and computationally difficult. In addition to its relevance as a typical model to formulate a number of practical problems, MSCP can be used as a benchmark problem to test constraint satisfaction algorithms and solvers.

Based on this review, we discuss some perspective research directions.
\begin{itemize}
\item \textit{Evaluation function and search space:} as introduced in Section \ref{sec_Definitions}, the aim of MSCP is twofold: (1) find a \emph{proper} $k$-coloring $c$ of a graph and (2) ensure that the sum of the colors assigned to the vertices is \emph{minimized}. An evaluation function combining these two objectives has been proposed in \cite{Helmar&Chiarandini2011}:
\begin{equation*}
f'(c)= \sum_{l=1}^k l|V_l| + M |E(V_l)|
\end{equation*}
where $E(V_l)$ is the set of conflicting edges in $V_l$ and $M > 0$ is a sufficiently large natural number. Since the evaluation function is used to guide the heuristic search process, it would be interesting to design other effective evaluation function based on a better recombination of the two parts of $f'$.

Another possibility could be to explore only the feasible graph coloring search space, like in the competitive MASC and MA-MSCP approaches \cite{Jin&al2014,Moukrim&al2013}, using more effective (multi-)neighborhood structures.

Besides, the combination of the above two ingredients in a proper way may lead to improved MSCP algorithms.

\item \textit{Maximum independent sets extraction:} As shown in Section \ref{subsec_greedyAlgorithms}, EXSCOL is a greedy heuristic based on the independent sets extraction that is quite effective for large graphs. Its major deficiency is that it does not include a procedure to reconsider ``bad'' independent sets that has been extracted. Hence, one possibility is to devise a backtracking procedure when a ``bad'' independent set has been identified as proposed for the graph coloring problem \cite{Wu&Hao2012b}.

\item \textit{Exact algorithms:} There is no exact algorithm especially designed for MSCP except the general approach which applies CPLEX to solve the integer linear programming formulation of MSCP \cite{Wang&al2013}.  However, as shown in \cite{Wang&al2013}, this approach is only applicable to easy DIMACS instances. On the other hand, some exact algorithms for the classical vertex coloring problem successfully solved a subset of the hard DIMACS graphs. Hence, it would be important to fill the gap by designing exact algorithms for MSCP.

\end{itemize}

To conclude, the minimum sum coloring problem, like the classical coloring problem, is a generic and useful model. Advances in solution methods (both exact and heuristic methods) for these coloring problems will help find satisfying solutions to many practical problems. Given the increasing interest in the sum coloring problem and their related coloring problems, it is reasonable to believe that research in these domains will become even more intense and fruitful in the forthcoming years.

\section*{Acknowledgment}
We are grateful to the anonymous referees for valuable suggestions and comments which have helped us to improve the paper. This work was partially supported by the LigeRO (2009-2014, Pays de la Loire Region), PGMO (2014-2016, Jacques Hadamard Mathematical Foundation) projects and the National Natural Science Foundation Program of China (Grant No. 61472147).

\newpage

\section*{Appendix}

For the purpose of completeness, this Appendix, which reproduces and extends the results given in \cite{Jin&Hao2016}, shows a performance summary of the six main heuristic algorithms for the set of 94 DIMACS benchmark graphs in terms of the lower and upper bounds of the MSCP problem.

\begin{landscape}
\begin{scriptsize}
\begin{longtable}{@{}l@{}r@{ }r@{ }c@{ }r@{ }r@{ }c@{ }r@{ }r@{ }c@{ }r@{ }r@{ }c@{ }r@{ }r@{ }r@{ }r@{ }c@{ }r@{ }r@{ }r@{ }r@{ }c@{ }r@{ }r@{ }r@{ }r@{}}
\caption{The performance of six heuristics and metaheuristics for the lower and upper bounds of MSCP} \label{table_many_algorithms}\\
\hline
\multicolumn{3}{c}{\multirow{3}{*}{Graph}} &&\multicolumn{2}{c}{BLS \cite{Benlic&Hao2012}} &&\multicolumn{2}{c}{MASC \cite{Jin&al2014}} &&\multicolumn{2}{c}{MDS(5)+LS \cite{Helmar&Chiarandini2011}} &&\multicolumn{4}{c}{EXSCOL \cite{Wu&Hao2012,Wu&Hao2013}} &&\multicolumn{4}{c}{MA-MSCP \cite{Moukrim&al2013}}  &&\multicolumn{4}{c}{HESA \cite{Jin&Hao2016}}\\
&&&&\multicolumn{2}{c}{2.83 GHz} &&\multicolumn{2}{c}{2.7 GHz} &&\multicolumn{2}{c}{2.93 GHz} &&\multicolumn{4}{c}{2.8 GHz, 2.83 GHz} &&\multicolumn{4}{c}{1.66 GHz}  &&\multicolumn{4}{c}{2.83 GHz}\\
&&&&\multicolumn{2}{c}{2 hours} &&\multicolumn{2}{c}{50 generations} &&\multicolumn{2}{c}{1 hour} &&\multicolumn{4}{c}{No stop condition} &&\multicolumn{4}{c}{2 hours}  &&\multicolumn{4}{c}{2 hours}\\
\cline{1-3}\cline{5-6}\cline{8-9}\cline{11-12}\cline{14-17}\cline{19-22}\cline{24-27}
Name&$f^b_{LB}$&$f^b_{UB}$&&$f^*_{UB}$&$f^a_{UB}$&&$f^*_{UB}$&$f^a_{UB}$&&$f^*_{LB}$&$f^*_{UB}$&&$f^*_{LB}$&$f^a_{LB}$&$f^*_{UB}$&$f^a_{UB}$&&$f^*_{LB}$&$f^a_{LB}$&$f^*_{UB}$&$f^a_{UB}$&&$f^*_{LB}$&$f^a_{LB}$&$f^*_{UB}$&$f^a_{UB}$\\
\hline
\endfirsthead
\multicolumn{27}{c}{\tablename\ \thetable{} Continued from previous page}\\
\hline
\multicolumn{3}{c}{\multirow{3}{*}{Graph}} &&\multicolumn{2}{c}{BLS \cite{Benlic&Hao2012}} &&\multicolumn{2}{c}{MASC \cite{Jin&al2014}} &&\multicolumn{2}{c}{MDS(5)+LS \cite{Helmar&Chiarandini2011}} &&\multicolumn{4}{c}{EXSCOL \cite{Wu&Hao2012,Wu&Hao2013}} &&\multicolumn{4}{c}{MA-MSCP \cite{Moukrim&al2013}}  &&\multicolumn{4}{c}{HESA \cite{Jin&Hao2016}}\\
&&&&\multicolumn{2}{c}{2.83 GHz} &&\multicolumn{2}{c}{2.7 GHz} &&\multicolumn{2}{c}{2.93 GHz} &&\multicolumn{4}{c}{2.8 GHz, 2.83 GHz} &&\multicolumn{4}{c}{1.66 GHz}  &&\multicolumn{4}{c}{2.83 GHz}\\
&&&&\multicolumn{2}{c}{2 hours} &&\multicolumn{2}{c}{50 generations} &&\multicolumn{2}{c}{1 hour} &&\multicolumn{4}{c}{No stop condition} &&\multicolumn{4}{c}{2 hours}  &&\multicolumn{4}{c}{2 hours}\\
\cline{1-3}\cline{5-6}\cline{8-9}\cline{11-12}\cline{14-17}\cline{19-22}\cline{24-27}
Name&$f^b_{LB}$&$f^b_{UB}$&&$f^*_{UB}$&$f^a_{UB}$&&$f^*_{UB}$&$f^a_{UB}$&&$f^*_{LB}$&$f^*_{UB}$&&$f^*_{LB}$&$f^a_{LB}$&$f^*_{UB}$&$f^a_{UB}$&&$f^*_{LB}$&$f^a_{LB}$&$f^*_{UB}$&$f^a_{UB}$&&$f^*_{LB}$&$f^a_{LB}$&$f^*_{UB}$&$f^a_{UB}$\\
\hline
\endhead
\hline \multicolumn{27}{c}{{Continued on next page}}
\endfoot
\hline
\endlastfoot
myciel3&16&21&&21&21.0&&21&21.0&&16&21&&16&16.0&21&21.0&&16&16.0&21&21.0&&16&16.0&21&21.0\\
myciel4&34&45&&45&45.0&&45&45.0&&34&45&&34&34.0&45&45.0&&34&34.0&45&45.0&&34&34.0&45&45.0\\
myciel5&70&93&&93&93.0&&93&93.0&&70&93&&70&70.0&93&93.0&&70&70.0&93&93.0&&70&70.0&93&93.0\\
myciel6&142&189&&189&196.6&&189&189.0&&142&189&&142&142.0&189&189.0&&142&139.5&189&189.0&&142&142.0&189&189.0\\
myciel7&286&381&&381&393.8&&381&381.0&&286&381&&286&286.0&381&381.0&&286&277.5&381&381.0&&286&286.0&381&381.0\\
anna&273&276&&276&276.0&&276&276.0&&273&276&&273&273.0&283&283.2&&273&273.0&276&276.0&&273&273.0&276&276.0\\
david&234&237&&237&237.0&&237&237.0&&234&237&&229&229.0&237&238.1&&234&234.0&237&237.0&&234&234.0&237&237.0\\
huck&\textbf{243}&\textbf{243}&&243&243.0&&243&243.0&&243&243&&243&243.0&243&243.8&&243&243.0&243&243.0&&243&243.0&243&243.0\\
jean&216&217&&217&217.0&&217&217.0&&216&217&&216&216.0&217&217.3&&216&216.0&217&217.0&&216&216.0&217&217.0\\
homer&1129&1150&&-&-&&1155&1158.5&&-&-&&-&-&-&-&&1129&1129.0&1157&1481.9&&1129&1129.0&1150&1151.8\\
queen5.5&\textbf{75}&\textbf{\underline{75}}&&75&75.0&&75&75.0&&75&75&&75&75.0&75&75.0&&75&75.0&75&75.0&&75&75.0&75&75.0\\
queen6.6&126&138&&138&138.0&&138&138.0&&126&138&&126&126.0&150&150.0&&126&126.0&138&138.0&&126&126.0&138&138.0\\
queen7.7&\textbf{196}&\textbf{\underline{196}}&&196&196.0&&196&196.0&&196&196&&196&196.0&196&196.0&&196&196.0&196&196.0&&196&196.0&196&196.0\\
queen8.8&288&291&&291&291.0&&291&291.0&&288&291&&288&288.0&291&291.0&&288&288.0&291&291.0&&288&288.0&291&291.0\\
queen8.12&\textbf{624}&\textbf{\underline{624}}&&-&-&&624&624.0&&-&-&&-&-&-&-&&624&624.0&624&624.0&&624&624.0&624&624.0\\
queen9.9&405&409&&-&-&&409&410.5&&-&-&&-&-&-&-&&405&405.0&409&411.9&&405&405.0&409&409.0\\
queen10.10&550&553&&-&-&&-&-&&-&-&&-&-&-&-&&550&550.0&553&555.2&&550&550.0&553&553.6\\
queen11.11&726&733&&-&-&&-&-&&-&-&&-&-&-&-&&726&726.0&733&735.4&&726&726.0&733&734.4\\
queen12.12&936&943&&-&-&&-&-&&-&-&&-&-&-&-&&936&936.0&944&948.7&&936&936.0&943&947.0\\
queen13.13&1183&1191&&-&-&&-&-&&-&-&&-&-&-&-&&1183&1183.0&1192&1197.0&&1183&1183.0&1191&1195.4\\
queen14.14&1470&1482&&-&-&&-&-&&-&-&&-&-&-&-&&1470&1470.0&1482&1490.8&&1470&1470.0&1482&1487.3\\
queen15.15&1800&1814&&-&-&&-&-&&-&-&&-&-&-&-&&1800&1800.0&1814&1823.0&&1800&1800.0&1814&1820.1\\
queen16.16&2176&2193&&-&-&&-&-&&-&-&&-&-&-&-&&2176&2176.0&2197&2205.9&&2176&2176.0&2193&2199.4\\
school1&2439&2674&&-&-&&-&-&&-&-&&-&-&-&-&&2345&2283.3&2674&2766.8&&2439&2418.9&2674&2674.0\\
school1-nsh&2176&2392&&-&-&&-&-&&-&-&&-&-&-&-&&2106&2064.6&2392&2477.1&&2176&2169.4&2392&2392.0\\
miles250&318&325&&327&328.8&&325&325.0&&318&325&&318&316.2&328&333.0&&318&318.0&325&325.4&&318&318.0&325&325.0\\
miles500&686&705&&710&713.3&&705&705.0&&686&712&&677&671.4&709&714.5&&686&686.0&708&711.2&&686&686.0&705&705.8\\
miles750&1145&1173&&-&-&&-&-&&-&-&&-&-&-&-&&1145&1145.0&1173&1183.9&&1145&1145.0&1173&1173.6\\
miles1000&1623&1666&&-&-&&-&-&&-&-&&-&-&-&-&&1623&1623.0&1679&1697.3&&1623&1623.0&1666&1670.5\\
miles1500&3239&3354&&-&-&&-&-&&-&-&&-&-&-&-&&3239&3239.0&3354&3357.2&&3239&3239.0&3354&3354.0\\
fpsol2.i.1&\textbf{3403}&\textbf{3403}&&-&-&&3403&3403.0&&3151&3403&&3403&3403.0&-&-&&3403&3403.0&3403&3403.0&&3403&3403.0&3403&3403.0\\
fpsol2.i.2&\textbf{1668}&\textbf{1668}&&-&-&&1668&1668.0&&-&-&&-&-&-&-&&1668&1668.0&1668&1668.0&&1668&1668.0&1668&1668.0\\
fpsol2.i.3&\textbf{1636}&\textbf{1636}&&-&-&&1636&1636.0&&-&-&&-&-&-&-&&1636&1636.0&1636&1636.0&&1636&1636.0&1636&1636.0\\
mug88\_1&164&178&&-&-&&178&178.0&&164&178&&164&162.3&-&-&&-&-&-&-&&164&164.0&178&178.0\\
mug88\_25&162&178&&-&-&&178&178.0&&162&178&&162&160.3&-&-&&-&-&-&-&&162&162.0&178&178.0\\
mug100\_1&188&202&&-&-&&202&202.0&&188&202&&188&188.0&-&-&&-&-&-&-&&188&188.0&202&202.0\\
mug100\_25&186&202&&-&-&&202&202.0&&186&202&&186&183.4&-&-&&-&-&-&-&&186&186.0&202&202.0\\
2-Insert\_3&55&62&&-&-&&62&62.0&&55&62&&55&55.0&-&-&&-&-&-&-&&55&55.0&62&62.0\\
3-Insert\_3&84&92&&-&-&&92&92.0&&84&92&&84&82.8&-&-&&-&-&-&-&&84&84.0&92&92.0\\
inithx.i.1&\textbf{3676}&\textbf{3676}&&-&-&&3676&3676.0&&3486&3676&&3676&3676.0&-&-&&3676&3616.0&3676&3679.6&&3676&3675.3&3676&3676.0\\
inithx.i.2&\textbf{2050}&\textbf{2050}&&-&-&&2050&2050.0&&-&-&&-&-&-&-&&2050&1989.2&2050&2053.7&&2050&2050.0&2050&2050.0\\
inithx.i.3&\textbf{1986}&\textbf{1986}&&-&-&&1986&1986.0&&-&-&&-&-&-&-&&1986&1961.8&1986&1986.0&&1986&1986.0&1986&1986.0\\
mulsol.i.1&\textbf{1957}&\textbf{1957}&&-&-&&1957&1957.0&&-&-&&-&-&-&-&&1957&1957.0&1957&1957.0&&1957&1957.0&1957&1957.0\\
mulsol.i.2&\textbf{1191}&\textbf{1191}&&-&-&&1191&1191.0&&-&-&&-&-&-&-&&1191&1191.0&1191&1191.0&&1191&1191.0&1191&1191.0\\
mulsol.i.3&\textbf{1187}&\textbf{1187}&&-&-&&1187&1187.0&&-&-&&-&-&-&-&&1187&1187.0&1187&1187.0&&1187&1187.0&1187&1187.0\\
mulsol.i.4&\textbf{1189}&\textbf{1189}&&-&-&&1189&1189.0&&-&-&&-&-&-&-&&1189&1189.0&1189&1189.0&&1189&1189.0&1189&1189.0\\
mulsol.i.5&\textbf{1160}&\textbf{1160}&&-&-&&1160&1160.0&&-&-&&-&-&-&-&&1160&1160.0&1160&1160.0&&1160&1160.0&1160&1160.0\\
zeroin.i.1&\textbf{1822}&\textbf{1822}&&-&-&&1822&1822.0&&-&-&&-&-&-&-&&1822&1822.0&1822&1822.0&&1822&1822.0&1822&1822.0\\
zeroin.i.2&\textbf{1004}&\textbf{1004}&&-&-&&1004&1004.0&&1004&1004&&1004&1004.0&-&-&&1004&1002.1&1004&1004.0&&1004&1004.0&1004&1004.0\\
zeroin.i.3&\textbf{998}&\textbf{998}&&-&-&&998&998.0&&998&998&&998&998.0&-&-&&998&998.0&998&998.0&&998&998.0&998&998.0\\
wap05&12449&13656&&-&-&&13669&13677.8&&-&-&&12428&12339.3&13680&13718.4&&-&-&-&-&&12449&12438.9&13656&13677.8\\
wap06&12454&13773&&-&-&&13776&13777.8&&-&-&&12393&12348.8&13778&13830.9&&-&-&-&-&&12454&12431.6&13773&13777.6\\
wap07&24800&28617&&-&-&&28617&28624.7&&-&-&&24339&24263.8&28629&28663.8&&-&-&-&-&&24800&24783.6&29154&29261.1\\
wap08&25283&28885&&-&-&&28885&28890.9&&-&-&&24791&24681.1&28896&28946.0&&-&-&-&-&&25283&25263.4&29460&29542.3\\
qg.order30&\textbf{13950}&\textbf{\underline{13950}}&&-&-&&13950&13950.0&&-&-&&13950&13950.0&13950&13950.0&&13950&13950.0&13950&13950.0&&13950&13950.0&13950&13950.0\\
qg.order40&\textbf{32800}&\textbf{\underline{32800}}&&-&-&&32800&32800.0&&-&-&&32800&32800.0&32800&32800.0&&32800&32800.0&32800&32800.0&&32800&32800.0&32800&32800.0\\
qg.order60&\textbf{109800}&\textbf{\underline{109800}}&&-&-&&109800&109800.0&&-&-&&109800&109800.0&110925&110993.0&&109800&109800.0&109800&109800.0&&109800&109800.0&109800&109800.0\\
DSJC125.1&247&326&&326&326.9&&326&326.6&&238&326&&246&244.1&326&326.7&&247&244.6&326&327.3&&247&247.0&326&326.1\\
DSJC125.5&549&1012&&1012&1012.9&&1012&1020.0&&493&1015&&536&522.4&1017&1019.7&&549&541.0&1013&1018.5&&549&548.5&1012&1012.2\\
DSJC125.9&1691&2503&&2503&2503.0&&2503&2508.0&&1621&2511&&1664&1592.5&2512&2512.0&&1689&1677.7&2503&2519.0&&1691&1691.0&2503&2503.0\\
DSJC250.1&570&970&&973&982.5&&974&990.5&&521&977&&567&562.0&985&985.0&&569&558.4&983&995.8&&570&569.2&970&980.4\\
DSJC250.5&1287&3210&&3219&3248.5&&3230&3253.7&&1128&3281&&1270&1258.8&3246&3253.9&&1280&1249.4&3214&3285.5&&1287&1271.6&3210&3235.6\\
DSJC250.9&4311&8277&&8290&8316.0&&8280&8322.7&&3779&8412&&4179&4082.4&8286&8288.8&&4279&4160.9&8277&8348.8&&4311&4279.4&8277&8277.2\\
DSJC500.1&1250&2836&&2882&2942.9&&2841&2844.1&&1143&2951&&1250&1246.6&2850&2857.4&&1241&1214.9&2897&2990.5&&1250&1243.4&2836&2836.0\\
DSJC500.5&2923&10886&&11187&11326.3&&10897&10905.8&&2565&11717&&2921&2902.6&10910&10918.2&&2868&2797.7&11082&11398.3&&2923&2896.0&10886&10891.5\\
DSJC500.9&11053&29862&&30097&30259.2&&29896&29907.8&&9731&30872&&10881&10734.5&29912&29936.2&&10759&10443.8&29995&30361.9&&11053&10950.1&29862&29874.3\\
DSJC1000.1&2762&8991&&9520&9630.1&&8995&9000.5&&2456&10123&&2762&2758.6&9003&9017.9&&2707&2651.2&9188&9667.1&&2719&2707.6&8991&8996.5\\
DSJC1000.5&6708&37575&&40661&41002.6&&37594&37597.6&&5660&43614&&6708&6665.9&37598&37673.8&&6534&6182.5&38421&40260.9&&6582&6541.3&37575&37594.7\\
DSJC1000.9&26557&103445&&-&-&&103464&103464.0&&23208&112749&&26557&26300.3&103464&103531.0&&26157&24572.0&105234&107349.0&&26296&26150.3&103445&103463.3\\
DSJR500.1&2069&2156&&-&-&&-&-&&-&-&&-&-&-&-&&2061&2052.9&2173&2253.1&&2069&2069.0&2156&2170.7\\
DSJR500.1c&15398&16286&&-&-&&-&-&&-&-&&-&-&-&-&&15025&14443.9&16311&16408.5&&15398&15212.4&16286&16286.0\\
DSJR500.5&22974&25440&&-&-&&-&-&&-&-&&-&-&-&-&&22728&22075.0&25630&26978.0&&22974&22656.7&25440&25684.1\\
flat300\_20\_0&1531&\underline{3150}&&-&-&&3150&3150.0&&-&-&&1524&1505.7&3150&3150.0&&1515&1479.3&3150&3150.0&&1531&1518.2&3150&3150.0\\
flat300\_26\_0&1548&3966&&-&-&&3966&3966.0&&-&-&&1525&1511.4&3966&3966.0&&1536&1501.6&3966&3966.0&&1548&1530.3&3966&3966.0\\
flat300\_28\_0&1547&4238&&-&-&&4238&4313.4&&-&-&&1532&1515.3&4282&4286.1&&1541&1503.9&4261&4389.4&&1547&1536.5&4260&4290.0\\
flat1000\_50\_0&6601&\underline{25500}&&-&-&&25500&25500.0&&-&-&&6601&6571.8&25500&25500.0&&6433&6121.5&25500&25500.0&&6476&6452.1&25500&25500.0\\
flat1000\_60\_0&6640&30100&&-&-&&30100&30100.0&&-&-&&6640&6600.5&30100&30100.0&&6402&6047.7&30100&30100.0&&6491&6466.5&30100&30100.0\\
flat1000\_76\_0&6632&37164&&-&-&&37167&37167.0&&-&-&&6632&6583.2&37167&37213.2&&6330&6074.6&38213&39722.7&&6509&6482.8&37164&37165.9\\
le450\_5a&1193&\underline{1350}&&-&-&&1350&1350.0&&-&-&&-&-&-&-&&1190&1171.5&1350&1350.0&&1193&1191.5&1350&1350.0\\
le450\_5b&1189&\underline{1350}&&-&-&&1350&1350.0&&-&-&&-&-&-&-&&1186&1166.5&1350&1350.0&&1189&1185.0&1350&1350.1\\
le450\_5c&1278&\underline{1350}&&-&-&&1350&1350.0&&-&-&&-&-&-&-&&1272&1242.3&1350&1350.0&&1278&1270.4&1350&1350.0\\
le450\_5d&1282&\underline{1350}&&-&-&&1350&1350.0&&-&-&&-&-&-&-&&1269&1245.2&1350&1350.0&&1282&1274.2&1350&1350.0\\
le450\_15a&2331&2632&&-&-&&2706&2742.6&&-&-&&2329&2313.7&2632&2641.9&&2329&2324.3&2681&2733.1&&2331&2331.0&2634&2648.4\\
le450\_15b&2348&2632&&-&-&&2724&2756.2&&-&-&&2343&2315.7&2642&2643.4&&2348&2335.0&2690&2730.6&&2348&2348.0&2632&2656.5\\
le450\_15c&2610&3487&&-&-&&3491&3491.0&&-&-&&2591&2545.3&3866&3868.9&&2593&2569.1&3943&4048.4&&2610&2606.6&3487&3792.4\\
le450\_15d&2628&3505&&-&-&&3506&3511.8&&-&-&&2610&2572.4&3921&3928.5&&2622&2587.2&3926&4032.4&&2628&2627.1&3505&3883.1\\
le450\_25a&3003&3153&&-&-&&3166&3176.8&&-&-&&2997&2964.4&3153&3159.4&&3003&3000.4&3178&3204.3&&3003&3003.0&3157&3166.7\\
le450\_25b&3305&3365&&-&-&&3366&3375.1&&-&-&&3305&3304.1&3366&3371.9&&3305&3304.1&3379&3416.2&&3305&3305.0&3365&3375.2\\
le450\_25c&3657&4515&&-&-&&4700&4773.3&&-&-&&3619&3597.1&4515&4525.4&&3638&3617.0&4648&4700.7&&3657&3656.9&4553&4583.8\\
le450\_25d&3698&4544&&-&-&&4722&4805.7&&-&-&&3684&3627.4&4544&4550.0&&3697&3683.2&4696&4740.3&&3698&3698.0&4569&4607.6\\
latin\_sqr\_10&40950&41444&&-&-&&41444&41481.5&&-&-&&40950&40950.0&42223&42392.7&&-&-&-&-&&40950&40950.0&41492&41672.8\\
C2000.5&15091&132483&&-&-&&-&-&&-&-&&15091&15077.6&132515&132682.0&&-&-&-&-&&14498&14442.9&132483&132513.9\\
C4000.5&33033&473234&&-&-&&-&-&&-&-&&33033&33018.8&473234&473211.0&&-&-&-&-&&31525&31413.3&513457&514639.0\\
games120&442&443&&443&443.0&&443&443.0&&442&443&&442&441.4&443&447.9&&442&442.0&443&443.0&&442&442.0&443&443.0\\
\end{longtable}
\end{scriptsize}
\end{landscape}

\end{document}